\documentclass{article}

\usepackage{PRIMEarxiv}

\usepackage[utf8]{inputenc} 
\usepackage[T1]{fontenc}    
\usepackage{hyperref}       
\usepackage{url}            
\usepackage{booktabs}       
\usepackage{amsfonts}       
\usepackage{nicefrac}       
\usepackage{microtype}      
\usepackage{lipsum}
\usepackage{fancyhdr}       
\usepackage{graphicx}       
\usepackage{amsmath} 
\graphicspath{{media/}}     

\title{TopoBind: Multi-Modal Prediction of Antibody-Antigen Binding Free Energy via Sequence Embeddings and Structural Topology
}

\author{
  Ciyuan Yu\thanks{\textbf{Equal contribution.}}, \\
  Department of Computer Science \\
  The City University of Hong Kong (Dongguan) \\
  Dongguan, China\\
  \texttt{72402757@cityu-dg.edu.cn} \\
  \And
  Hongzong Li$^*$, \\
  Generative AI Research and Development Center \\
  The Hong Kong University of Science and Technology \\
  Hong Kong\\
  \texttt{lihongzong@ust.hk} \\
   \And
   Jiahao Ma$^*$, \\
   Material Innovation Institute for Life Sciences and Energy\\
   The University of Hong Kong\\
   Hetao SZ-HK Cooperation Zone\\
  \texttt{jiahao.ma@connect.hku.hk} \\
   \And
  Shiqin Tang,\\
  Center for Artificial Intelligence and Robotics \\
  Chinese Academy of Science \\
  Hong Kong\\
  \texttt{shiqin.tang@cair-cas.org.hk} \\
   \And
  Ye-Fan Hu\thanks{
  \textbf{Corresponding authors.}}\,\,,\\
  Computational Immunology Centre \\
  BayVax Biotech Limited \\
  Hong Kong\\
  \texttt{yefan.hu@bayvaxbio.com} \\  
   \And
  Jian-Dong Huang$^\dagger$\\
  School of Biomedical Sciences \\
  The University of Hong Kong \\
  Hong Kong\\
  \texttt{jdhuang@hku.hk} \\
}

\begin{document}
\maketitle


\begin{abstract}
Predicting the binding free energy between antibodies and antigens is a key challenge in structure-aware biomolecular modeling, with direct implications for antibody design. Most existing methods either rely solely on sequence embeddings or struggle to capture complex structural relationships, thus limiting predictive performance. In this work, we present a novel framework that integrates sequence-based representations from pre-trained protein language models (ESM-2) with a set of topological features. Specifically, we extract contact map metrics reflecting residue-level connectivity, interface geometry descriptors characterizing cross-chain interactions, distance map statistics quantifying spatial organization, and persistent homology invariants that systematically capture the emergence and persistence of multi-scale topological structures—such as connected components, cycles, and cavities—within individual proteins and across the antibody-antigen interface. By leveraging a cross-attention mechanism to fuse these diverse modalities, our model effectively encodes both global and local structural organization, thereby substantially enhancing the prediction of binding free energy. Extensive experiments demonstrate that our model consistently outperforms sequence-only and conventional structural models, achieving state-of-the-art accuracy in binding free energy prediction. 

\end{abstract}


\section{Introduction}

Predicting the binding free energy ($\Delta G$) of antibody-antigen complexes is essential for antibody engineering and therapeutic design. Experimental methods for measuring $\Delta G$, such as surface plasmon resonance and isothermal titration calorimetry, are often costly and low-throughput. Computational estimation of $\Delta G$ thus plays a key role in accelerating antibody screening pipelines. However, traditional physics-based approaches such as molecular dynamics (MD) and MM/PBSA suffer from limited scalability and high computational cost \cite{wang2019accurate}.

To address these challenges, recent advances in protein language models (PLMs) have shown great promise. Models like ESM \cite{rives2021esm} and ProtTrans \cite{elnaggar2021prottrans} use transformer-based architectures pretrained on large-scale protein sequence data to generate embeddings that encode evolutionary and biochemical signals. These embeddings have achieved strong performance on a variety of downstream tasks, including contact map prediction, secondary structure classification, and protein-protein interaction detection \cite{rao2019evaluating, brandes2022proteinbert}. 

Nevertheless, PLMs inherently lack spatial awareness and may fail to capture the 3D structural characteristics that are critical for molecular recognition. Structure-informed models aim to bridge this gap by explicitly modeling geometric patterns through graph-based representations or surface-aware learning \cite{gainza2020deciphering, gligorijevic2021structure, townshend2021atom3d, ingraham2019generative}. However, many of these methods overlook global topological invariants—stable structural features that persist under deformation.

Persistent homology (PH), a key technique in topological data analysis (TDA), offers a principled way to extract such features from molecular structures \cite{cang2018representability}. By capturing connected components, loops, and voids across multiple filtration scales, PH enables robust, multiscale characterization of protein geometry. Prior studies have demonstrated the utility of PH-based features in function prediction and binding site identification \cite{pun2018persistent}.

Despite the complementarity of sequence and topological information, few existing models integrate both modalities for regression-based prediction of binding free energy ($\Delta G$). Most works focus on classification tasks such as interface detection \cite{chen2023residueattention} or docking \cite{townshend2021atom3d}, leaving continuous $\Delta G$ prediction largely underexplored.

In this work, we propose \textbf{TopoBind}, a multi-modal deep learning framework for predicting antibody-antigen binding free energy ($\Delta G$) by integrating pretrained sequence embeddings from ESM-2 with handcrafted structural topology features. Our model employs separate encoders for each modality, adaptively fuses four categories of geometric descriptors via a learnable gating module, and applies a bidirectional cross-attention mechanism to align sequence and structural representations. The fused embedding is further processed by either a multi-layer predictor or a sparse Lasso regressor.

Our key contributions are summarized as follows:
\begin{itemize}
    \item We develop a unified and interpretable multi-modal architecture that integrates pretrained sequence information with diverse geometric and topological representations of antibody-antigen interfaces.
    \item We design a rich 100-dimensional topological feature vector, consisting of contact statistics, interface geometry, distance-based descriptors, and persistent homology lifetimes, capturing both local and global structural characteristics.
    \item We introduce an adaptive feature fusion (AFF) mechanism to dynamically weight different topological submodules and demonstrate that it improves model generalization in conjunction with cross-attention and sparse regression.
    \item We validate our method on a curated dataset of 303 antibody-antigen complexes with experimentally measured $\Delta G$, achieving superior performance over classical baselines and strong neural models in both regression and classification settings.
\end{itemize}

\section{Related Work}
\subsection{Physics-Based Affinity Estimation}
Physics-based methods have long been used to estimate the binding affinity of protein complexes. Techniques such as molecular dynamics (MD), MM/PBSA, and MM/GBSA provide detailed physical insights but are computationally intensive and sensitive to conformational variability \cite{genheden2015mmgbsa}. This makes them less practical for large-scale or real-time applications in antibody discovery.
\subsection{Classical Machine Learning Models}
To improve scalability, classical machine learning models have been proposed as efficient alternatives. For example, COMBINE analysis uses regression on energy decomposition profiles \cite{ganotra2018combine}, while structure-based models like PDA-Pred and PRA-Pred incorporate handcrafted features such as interface area and hydrogen bonds to predict affinity in protein-DNA and protein-RNA systems \cite{pdapred2023, prapred2024}. However, these models typically depend on predefined descriptors and lack generalizability to diverse protein interfaces.
\subsection{Sequence and Structure-Based Deep Learning}
The advent of protein language models (PLMs) has introduced new possibilities for sequence-based binding affinity estimation. ESM-2 \cite{rives2021esm} and ProtTrans \cite{elnaggar2021prottrans} are pretrained on massive protein databases and can extract unsupervised embeddings that generalize across multiple tasks. These models have achieved state-of-the-art results in contact prediction and functional annotation \cite{rao2019evaluating, brandes2022proteinbert}, but often underperform in spatially-sensitive tasks such as residue-level interaction prediction \cite{cai2024gearbind}.

To address this, structure-informed deep learning frameworks have emerged. GearBind \cite{cai2024gearbind} and DeepPPAPred \cite{chakrabarty2025deepppapred} utilize 3D convolutional networks or graph-based transformers to model residue-level spatial arrangements, enhancing affinity prediction through geometric learning. FuncPhos-STR \cite{zhang2024funcphos} further integrates predicted structural dynamics from AlphaFold into site-level prediction, demonstrating the benefits of combining structure with deep learning.
\subsection{Topological Data Analysis in Protein Modeling}
Topological data analysis (TDA) offers an orthogonal perspective by modeling shape and connectivity using persistent homology. PH captures multi-scale topological features—such as loops and voids—that are stable under perturbations, and has been successfully applied in protein folding, molecular dynamics, and ligand binding \cite{cang2018representability}.However, its integration into antibody-antigen modeling remains limited, despite its potential to complement structure-based learning\cite{liu2022pareto}.

\section{Method}

We propose a multi-modal regression framework that integrates sequence-level representations and topological features for predicting the binding free energy $\Delta G$ of antibody-antigen complexes. This section presents the formal problem definition, the topological feature extraction process, and the regression model architecture.
\begin{figure}[htbp]
    \centering
    \includegraphics[width=0.6\linewidth]{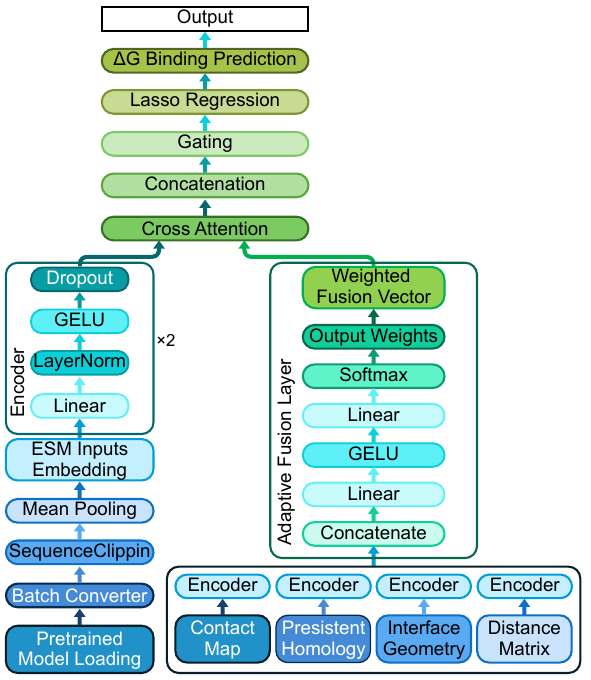}
    \caption{Overview of the TopoBind framework. The model integrates sequence and topological modalities for predicting antibody-antigen binding free energy. \textbf{Left:} ESM2-derived sequence embeddings are extracted via mean pooling and encoded using a two-layer MLP. \textbf{Right:} Structural topology is decomposed into four categories—contact maps, interface geometry, distance matrices, and persistent homology—which are individually encoded and fused via an adaptive weighting layer. \textbf{Top:} The resulting sequence and topology representations are aligned through two layers of bidirectional cross attention, then fused via a gating mechanism. The final embedding is passed to a Lasso regression module for $\Delta G$ prediction.}
    \label{fig:model}
\end{figure}

\subsection{Problem Formulation}

Let each antibody-antigen complex be represented by a protein sequence pair $(S_a, S_g)$ and its associated 3D structure. Our goal is to learn a mapping $f_\theta: \mathbb{R}^{d_s + d_t} \rightarrow \mathbb{R}$ that predicts the scalar binding free energy $\hat{y} = f_\theta(x)$, where $x = [x_\text{seq}, x_\text{topo}]$ is a concatenation of two feature vectors: one derived from pretrained sequence embeddings ($x_\text{seq} \in \mathbb{R}^{d_s}$), and one from handcrafted structural features ($x_\text{topo} \in \mathbb{R}^{d_t}$).

Given a training set $\{(x^{(i)}, y^{(i)})\}_{i=1}^N$ with experimentally measured $\Delta G$ values, we minimize the standard mean squared error:

\begin{equation}
\mathcal{L}(\theta) = \frac{1}{N} \sum_{i=1}^{N} \left( f_\theta(x^{(i)}) - y^{(i)} \right)^2
\label{eq:loss}
\end{equation}

This objective ensures that predicted binding free energy values remain close to the ground truth on average.

\subsection{Sequence Representation}

We leverage ESM-2, a pretrained transformer model, to encode the concatenated antibody and antigen sequence $S = [S_a ; S_g]$ into residue-wise hidden states $H \in \mathbb{R}^{L \times d}$. Here, $L$ is the total sequence length and $d$ is the embedding dimension (e.g., $d=2560$ for ESM2-3B). A mean pooling operation aggregates these residue-level embeddings into a global sequence descriptor:(see Fig.~\ref{fig:model})

\begin{equation}
x_\text{seq} = \frac{1}{L} \sum_{i=1}^{L} H_i
\end{equation}

This representation captures the evolutionary and contextual semantics of the entire complex.

\subsection{Topological Feature Extraction}

We extract $x_\text{topo}$ from four geometric representations of the antibody-antigen structure. These features capture multiscale interaction patterns from contact connectivity, interfacial geometry, Euclidean distances, and topological persistence(see Fig.~\ref{fig:model}).

\textbf{Contact Map Features.}  
We construct a binary matrix $C \in \{0,1\}^{n \times n}$ indicating whether a pair of residues is in contact (e.g., $C_{ij}=1$ if the $C_\alpha$-$C_\alpha$ distance is below threshold $\delta$). Based on this, we define:

\begin{equation}
D = \frac{1}{n^2} \sum_{i,j} C_{ij}
\end{equation}
\textit{Contact density}, measuring the overall sparsity of the contact network.

\begin{equation}
k_i = \sum_j C_{ij}, \quad \mu_k = \frac{1}{n} \sum_i k_i
\end{equation}
$k_i$ is the \textit{contact degree} of residue $i$, and $\mu_k$ is the average number of contacts per residue.

\begin{equation}
\sigma_k = \sqrt{ \frac{1}{n} \sum_i (k_i - \mu_k)^2 }, \quad k_{\max} = \max_i k_i
\end{equation}
Standard deviation $\sigma_k$ quantifies structural irregularity, and $k_{\max}$ identifies hotspots with maximal connectivity.

\begin{equation}
\bar{C} = \frac{ \text{Tr}(C^3) }{ \sum_i k_i(k_i - 1) }
\end{equation}
\textit{Clustering coefficient} $\bar{C}$ reflects the local density of triangles in the graph, indicating compactness and modularity.

\textbf{Interface Geometry Features.}  
For two chains (antibody and antigen), we define the binary interface contact matrix $I \in \{0,1\}^{n_a \times n_g}$. From this we compute:

\begin{equation}
D_I = \frac{1}{n_a n_g} \sum_{i,j} I_{ij}
\end{equation}
\textit{Interface density}, capturing how tightly the two proteins interact.

\begin{align}
R_a &= \frac{1}{n_a} \sum_i \mathbb{1} \left[ \sum_j I_{ij} > 0 \right], \\
R_g &= \frac{1}{n_g} \sum_j \mathbb{1} \left[ \sum_i I_{ij} > 0 \right]
\end{align}
\textit{Interface coverage}, measuring the proportion of residues involved in cross-chain contacts.

\begin{align}
\mu_a = \frac{1}{\sum_i \mathbb{1}[a_i > 0]} \sum_{i: a_i > 0} a_i, \\
\mu_g = \frac{1}{\sum_j \mathbb{1}[g_j > 0]} \sum_{j: g_j > 0} g_j
\end{align}
\textit{Mean interface degree}, assessing how many contacts each interacting residue participates in.

\textbf{Distance Map Features.}  
From the Euclidean distance matrix $D \in \mathbb{R}^{n \times n}$ between residues, we compute:

\begin{equation}
d_{\min} = \min_{i,j} D_{ij}, \quad \mu_D = \frac{1}{n^2} \sum_{i,j} D_{ij}
\end{equation}
\textit{Minimum} and \textit{mean internal distance} measure spatial compactness.

\begin{equation}
d_{\text{med}} = \text{median}( \{ D_{ij} \} )
\end{equation}
A robust estimator of internal packing.

\begin{equation}
\mu_I = \frac{1}{|D_I|} \sum_{(i,j) \in D_I} D_{ij}, \quad
N_c = |\{ (i,j) : D_{ij} \leq \delta \}|
\end{equation}
\textit{Interface mean distance} $\mu_I$ and \textit{contact count} $N_c$ summarize interaction range and density.

\textbf{Persistent Homology Features.}  
Persistent homology (PH) is a key method in topological data analysis (TDA), designed to quantify multiscale topological structures in geometric data. In our setting, PH captures the evolution of connected components, loops, and voids formed by atoms in 3D space, offering descriptors that are invariant to spatial transformations and robust to noise.

Given a point cloud $X = \{x_i \in \mathbb{R}^3\}_{i=1}^n$ extracted from atomic coordinates (e.g., heavy atoms or $C_\alpha$), we construct a Vietoris–Rips filtration: for a threshold $\epsilon > 0$, a simplex is added for each group of points with pairwise distance less than $\epsilon$. As $\epsilon$ increases, simplicial complexes $K_\epsilon$ are formed, and topological features appear (birth) and vanish (death).

Each topological feature corresponds to a persistence pair $(b_i^k, d_i^k)$ in dimension $k$. From these, we compute lifetimes $\ell_i = d_i^k - b_i^k$ and derive the following descriptors:

\begin{align}
N_k = |\{(b_i^k, d_i^k)\}|, \quad
\mu_k = \frac{1}{N_k} \sum_i \ell_i, \quad \\
\sigma_k = \sqrt{ \frac{1}{N_k} \sum_i (\ell_i - \mu_k)^2 }, 
\ell_{\max} = \max_i \ell_i
\end{align}

We also extract the top-$k$ lifetimes $\{\ell_{(1)}, \ldots, \ell_{(k)}\}$ to retain the most prominent features, sorted by persistence.These descriptors complement sequence-based features by providing robust summaries of global structural organization at the antibody-antigen interface.

\subsection{Regression Model}

Our regression framework integrates sequence and topological features through a multi-stage neural network (see Fig.~\ref{fig:model}).

\paragraph{Feature Encoders.}
We begin by encoding the two input modalities separately. The ESM-derived sequence embedding $x_\text{seq} \in \mathbb{R}^{d_s}$ is obtained by mean pooling the residue-level outputs from the pretrained ESM-2 model. This global representation is then passed through a two-layer feedforward encoder with LayerNorm, GELU activation, and Dropout:

\begin{equation}
h_\text{seq} = \mathrm{Block}_2^\mathrm{seq}(\mathrm{Block}1^\mathrm{seq}(x\text{seq}))
\end{equation}
\begin{equation}
\mathrm{Block}_i(z) = \mathrm{Dropout}(\mathrm{GELU} (\mathrm{LayerNorm}(\mathrm{Linear}i(z))))
\end{equation}
for $i \in {1,2}$, yielding the sequence encoding $h\text{seq} \in \mathbb{R}^{d_h}$.

For the topological input $x_\text{topo} \in \mathbb{R}^{d_t}$, we divide it into four semantically distinct sub-vectors:
\begin{itemize}
\item $x_c$: contact map features (e.g., density, degree statistics)
\item $x_i$: interface geometry features (e.g., coverage, cross-chain counts)
\item $x_d$: distance-based descriptors (e.g., mean/min distance)
\item $x_t$: persistent homology statistics (e.g., lifetime moments, top-$k$ persistence)
\end{itemize}

Each sub-vector is processed by an independent encoder:
\begin{align}
f_c &= \mathrm{Enc}_c(x_c), \quad f_i = \mathrm{Enc}_i(x_i) \\
f_d &= \mathrm{Enc}_d(x_d), \quad f_t = \mathrm{Enc}_t(x_t)
\end{align}
Each encoder follows the same architecture: two linear layers with GELU, LayerNorm, and Dropout. These four encoded vectors are then forwarded to the adaptive fusion stage.
\paragraph{Adaptive Feature Fusion (AFF).}
To synthesize information from the four topological subspaces, we introduce an Adaptive Feature Fusion (AFF) module, which computes attention-style weights to combine the encoded vectors into a single topology descriptor. Specifically:

\begin{align}
f_\text{concat} &= [f_c ; f_i ; f_d ; f_t], \quad f_\text{concat} \in \mathbb{R}^{4d_h} \\
w &= \mathrm{Softmax}(W_\text{gate} f_\text{concat} + b), \quad w \in \mathbb{R}^4 \\
h_\text{topo} &= w_1 \cdot f_c + w_2 \cdot f_i + w_3 \cdot f_d + w_4 \cdot f_t
\end{align}

Here, $W_\text{gate} \in \mathbb{R}^{4 \times 4d_h}$ and $b \in \mathbb{R}^4$ are trainable parameters. The softmax weights $w$ are dynamically adjusted for each sample, enabling data-dependent emphasis over different topology sources.

This adaptive fusion strategy enables the model to automatically prioritize the most informative geometric perspectives across examples, rather than treating all subspaces equally. The resulting $h_\text{topo} \in \mathbb{R}^{d_h}$ serves as the topology representation input to the cross-modal fusion layer.This entire module is visualized in the lower right section of Figure~\ref{fig:model}, emphasizing its central role in aggregating diverse geometric views.

\paragraph{Cross-Attention Fusion.}
We apply a two-layer bidirectional multi-head cross-attention module to model the interactions between sequence and topology representations. In each layer, the sequence embedding $h_\text{seq}$ attends to $h_\text{topo}$ and vice versa:

\begin{align}
h_\text{seq} &\leftarrow \mathrm{Norm}\left(h_\text{seq} + \mathrm{CrossAttn}(Q = h_\text{topo}, K = h_\text{seq})\right) \\
h_\text{topo} &\leftarrow \mathrm{Norm}\left(h_\text{topo} + \mathrm{CrossAttn}(Q = h_\text{seq}, K = h_\text{topo})\right) \\
h_\text{seq} &\leftarrow \mathrm{Norm}(h_\text{seq} + \mathrm{FFN}(h_\text{seq})) \\
h_\text{topo} &\leftarrow \mathrm{Norm}(h_\text{topo} + \mathrm{FFN}(h_\text{topo}))
\end{align}

Each attention layer uses $N_\text{head} = 8$ heads with hidden size $d_h = 256$. The two updated vectors are then prepared for downstream fusion and prediction.

\begin{figure}[ht]
    \centering
    \includegraphics[width=0.6\linewidth]{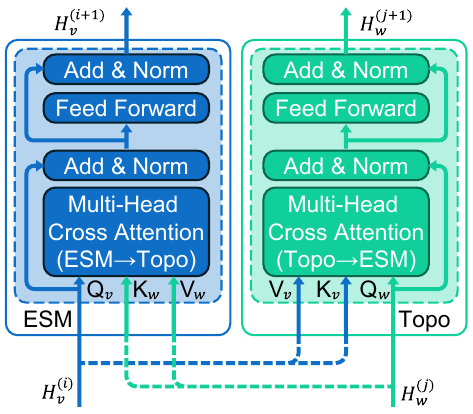}
    \caption{Multi-head cross attention fusion module.}
    \label{fig:cross_attention}
\end{figure}

\paragraph{Gated Feature Fusion.}
After cross-attention, we concatenate the updated sequence and topology embeddings, and apply a learnable gating mechanism to modulate their contribution. Specifically:
\begin{equation}
h_\text{fuse} = [h_\text{seq}; h_\text{topo}], \qquad
g = \sigma(W_g h_\text{fuse} + b_g)
\end{equation}
where $g \in [0,1]^{d_h}$ is a learnable gate vector that is a learnable gate vector that adaptively modulates the fused representation. The output is then passed to the downstream regression module.

\paragraph{Regression Head.}
The fused feature $h_\text{fuse}$ is input to a lasso regression model.Given a dataset of extracted fused representations $X \in \mathbb{R}^{N \times d}$ and ground-truth binding affinities $y \in \mathbb{R}^{N}$, the Lasso model learns a sparse linear predictor $\hat{y} = X w$ by minimizing the following regularized objective:
\begin{equation}
\min_{w \in \mathbb{R}^{d}} \frac{1}{2N} \| Xw - y \|_2^2 + \alpha \| w \|_1
\end{equation}
Here, $\alpha > 0$ controls the sparsity of the solution and is selected via cross-validation. This approach promotes sparsity, interpretability, and feature selection while leveraging the expressiveness of the upstream deep model.

\section{Experiments}

\subsection{Dataset}

Our dataset consists of 1705 antibody-antigen complexes from 472 unique PDB IDs. Among these, 472 samples include ESM-derived embeddings and 303 have computed topological features. A total of 303 PDB IDs with both modalities yield 1398 usable complex instances.Each instance is paired with an experimental binding free energy, forming a regression task. The data are split into 978 training, 209 validation, and 211 test samples, ensuring no PDB ID overlap across splits.
ESM-2 embeddings are mean-pooled into 2560-dimensional vectors. Topological features—spanning contact maps, interface geometry, distance matrices, and persistent homology—are padded or truncated to 100 dimensions based on the 90th percentile.

\subsection{Implementation Details}

The model encodes the concatenated antibody-antigen sequence using the pretrained ESM-2 (3B) model, followed by mean pooling to obtain a 2560-dimensional sequence embedding. Meanwhile, the topological feature vector is divided into four categories—contact statistics, interface geometry, distance metrics, and persistent homology descriptors—each processed by a dedicated two-layer feedforward encoder with LayerNorm, GELU, and Dropout. The resulting sub-embeddings are adaptively fused via a learnable gating mechanism to form a 256-dimensional topological representation.

The sequence and topology embeddings are aligned through two stacked bidirectional cross-attention layers (8 heads, hidden dimension 256), then fused via a sigmoid-based gating module. The final representation is input to a Lasso regressor to predict binding free energy ($\Delta G$).

We use the AdamW optimizer with an initial learning rate of 3e-4, weight decay of 1e-4, and a batch size of 32. ReduceLROnPlateau is used for learning rate scheduling (patience 7), and early stopping is triggered after 15 epochs without validation improvement. All experiments are conducted on a single NVIDIA A100 GPU with mixed-precision training.

\subsection{Baseline Methods}

We evaluate the proposed TopoBind (LASSO) model against several representative baselines using the same training and evaluation splits. These include an ESM2-only MLP model, a Random Forest regressor with PCA and RFECV-based feature selection, and the COMBINE method. In addition, we conduct ablation studies to assess the impact of the AFF module and compare LASSO- and MLP-based regressors within the TopoBind framework.

\paragraph{ESM2-only MLP.} A two-layer multilayer perceptron trained on the 2560-dimensional ESM embeddings without any structural information. This baseline follows the strategy of direct sequence-based regression using pretrained language model representations \cite{lin2022evolutionary}.

\paragraph{Random Forest with RFECV.}
A classical ensemble baseline using only ESM-derived sequence embeddings. The 2560-dimensional features are first reduced via PCA to preserve 95\% of the variance, followed by recursive feature elimination with cross-validation (RFECV) for supervised selection. The selected components are used to train a Random Forest Regressor with 200 trees and a maximum depth of 10. \cite{sahoo2025nanobep}.

\paragraph{COMBINE.} A classical regression approach based on energy decomposition and principal component regression. Interaction energy terms derived from molecular mechanics calculations are combined into a feature vector, which is then fitted with Partial Least Squares (PLS) regression. This model is included to represent energy-based scoring baselines commonly used in virtual screening pipelines.\cite{schapira1999prediction}

Table~\ref{tab:baseline-results} summarizes the test performance of all models across six evaluation metrics. The proposed TopoBind (LASSO) achieves the best results in all metrics, including the lowest MSE (3.8160) and highest $R^2$ (0.3390), confirming its superior regression capability over both neural and traditional baselines.

\begin{table*}[ht]
\centering
\caption{Performance comparison of different methods. Best results are highlighted in \textbf{bold} and second-best are \underline{underlined}.}
\label{tab:baseline-results}
\begin{tabular}{lcccccc}
\hline
\textbf{Model} & \textbf{MSE} $\downarrow$ & \textbf{RMSE} $\downarrow$ & \textbf{MAE} $\downarrow$ & \textbf{Correlation} $\uparrow$ & \textbf{$R^2$} $\uparrow$ & \textbf{Accuracy} $\uparrow$ \\
\hline
\textbf{TopoBind(LASSO)} & \textbf{ 3.8160} & \textbf{ 1.9535} & \textbf{1.4255} & \textbf{0.5905} & \textbf{0.3390} & \textbf{0.7500} \\
\underline{TopoBind (MLP)} & \underline{3.8965} & \underline{1.9739} & \underline{1.4499} & \underline{0.5754} & \underline{0.3251} & \underline{0.7214} \\
ESM2 & 4.5654 & 2.1367& 1.6288  & 0.5348 &  0.2679 &0.6569  \\
Random Forest + RFECV & 4.8176 & 2.1949 &  1.6008 & 0.5175 & 0.2359& 0.6758\\
COMBINE & 6.0487 & 2.4594 & 1.8805 & 0.0838 & 0.0058 & 0.6019 \\
\hline
\end{tabular}
\end{table*}

\begin{figure}[ht]
\centering
\includegraphics[width=0.6\columnwidth]{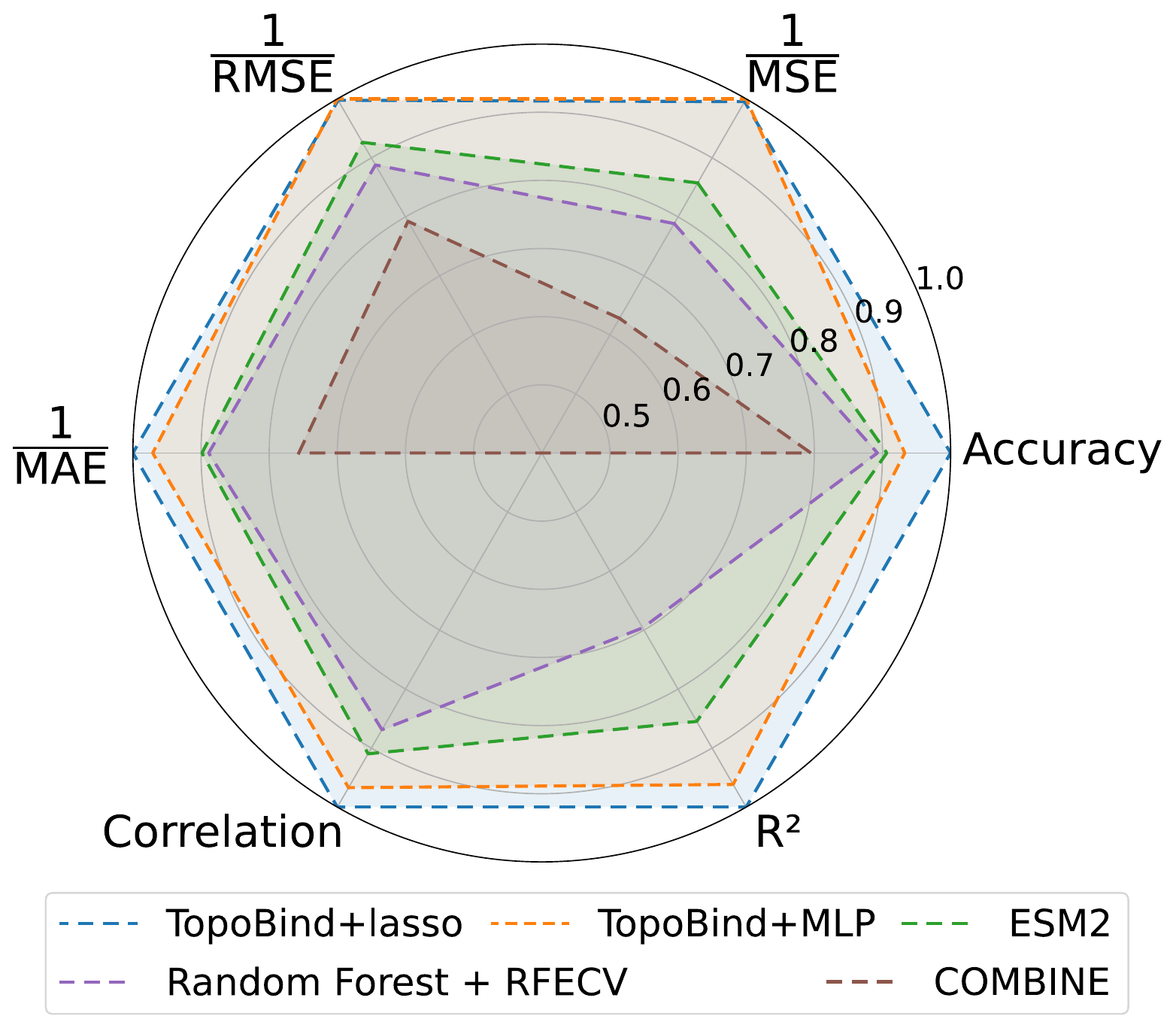}
\caption{Radar chart comparing the performance of different models across six evaluation metrics: MSE, RMSE, MAE, Pearson correlation, $R^2$, and classification accuracy. To ensure visual consistency, MSE, RMSE, and MAE are plotted using their reciprocals, so that larger values uniformly indicate better performance. A larger enclosed area thus reflects more favorable and well-rounded predictive capability. TopoBind (LASSO) shows the most balanced and expanded coverage.}
\label{fig:radar_comparison}
\end{figure}

Figure~\ref{fig:radar_comparison} visualizes the performance distribution of selected models across six evaluation metrics. To unify the interpretation direction, we apply reciprocal transformation to the three error-based metrics (MSE, RMSE, MAE), such that higher values consistently indicate better performance across all dimensions. TopoBind (LASSO) demonstrates superior and balanced results, forming the most expansive and regular radar shape. This highlights the model’s ability to maintain strong generalization without overfitting to any single objective.

\begin{figure}[ht]
\centering
\includegraphics[width=0.6\linewidth]{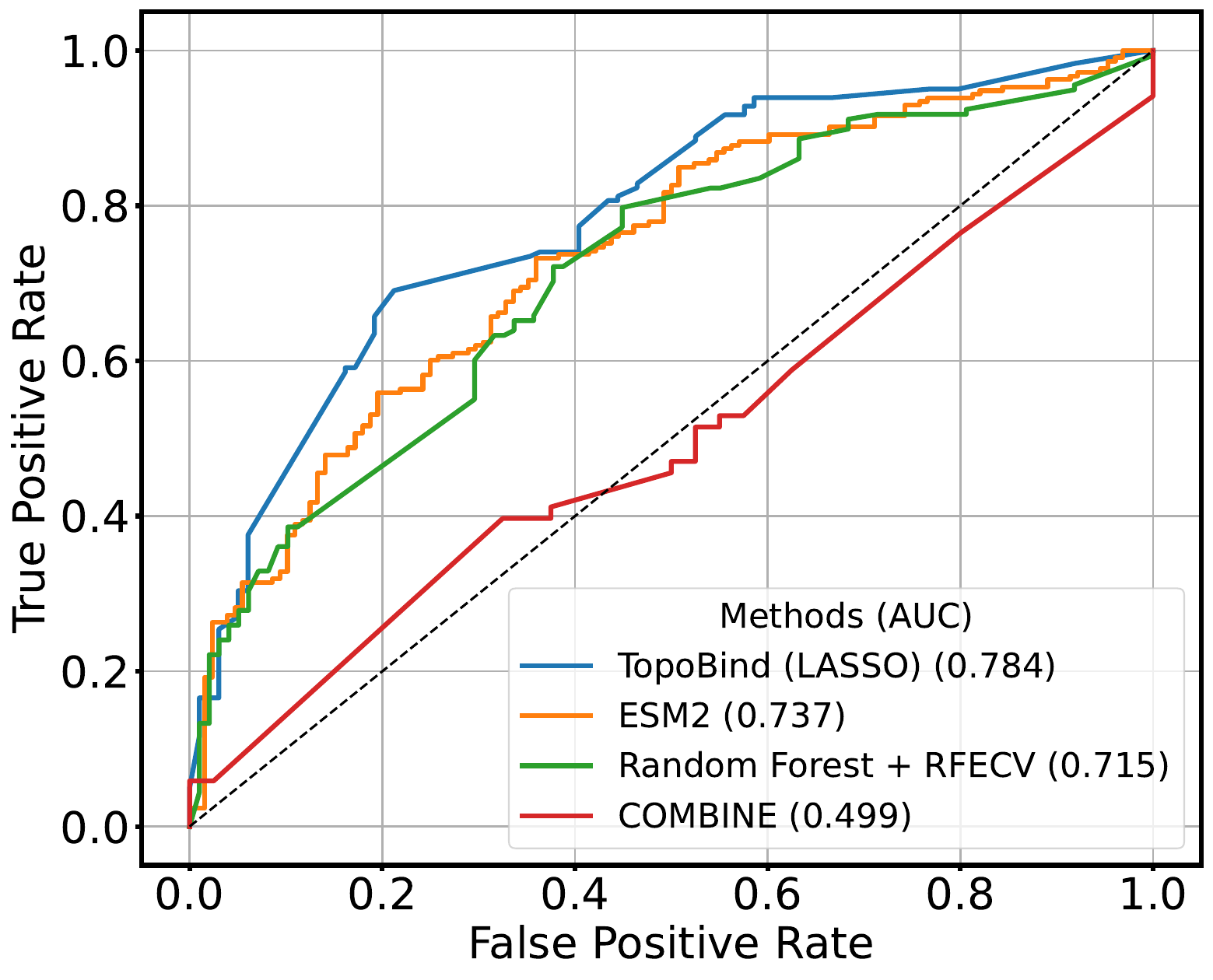}
\caption{Receiver Operating Characteristic (ROC) curves for five representative models. TopoBind (LASSO) achieves the highest area under the curve (AUC), indicating superior discriminative ability. In contrast, COMBINE approaches random performance, highlighting the limitations of classical energy-based scoring methods for antibody-antigen affinity prediction.}
\label{fig:roc_overview}
\end{figure}

Figure~\ref{fig:roc_overview} presents ROC curves for five representative models in classifying strong binders ($\Delta G < -10$). TopoBind (LASSO) achieves the highest AUC, reflecting its superior discriminative capability. In contrast, the COMBINE method performs comparably to random guessing, underscoring the limitations of classical energy-based scoring when applied to antibody-antigen systems. Our model benefits from deep cross-modal alignment and sparse prediction, enabling more reliable classification of biologically meaningful binding events.

\subsection{Ablation Study.}
To evaluate the importance of key architectural components, we conduct a series of ablation experiments summarized in Table~\ref{tab:ablation-results}. The full TopoBind model with both Adaptive Feature Fusion (AFF), Cross-Attention, and LASSO regression achieves the best $R^2$ score of 0.3390. Removing the AFF module causes a performance drop to 0.3201, indicating that dynamically weighting topological submodules improves generalization by enhancing feature complementarity. Eliminating the cross-attention module while keeping both input modalities leads to a further decline ($R^2$ = 0.2716), suggesting that explicit cross-modal alignment is crucial for integrating sequence and structural representations. Replacing the LASSO regression head with a two-layer MLP results in an $R^2$ of 0.3251, highlighting the role of sparse regression in mitigating overfitting and enhancing robustness. As a unimodal baseline, ESM2-only (LASSO) achieves a lower $R^2$ of 0.2896, confirming the benefit of incorporating geometric and topological priors. These findings collectively demonstrate that each architectural component of TopoBind contributes meaningfully to overall performance.

\begin{table*}[ht]
\centering
\caption{Ablation study of TopoBind architecture. Best results are highlighted in \textbf{bold} and second-best are \underline{underlined}.}
\label{tab:ablation-results}
\begin{tabular}{lcccccc}
\hline
\textbf{Model} & \textbf{MSE} $\downarrow$ & \textbf{RMSE} $\downarrow$ & \textbf{MAE} $\downarrow$ & \textbf{Correlation} $\uparrow$ & \textbf{$R^2$} $\uparrow$ & \textbf{Accuracy} $\uparrow$ \\
\hline
\textbf{TopoBind (LASSO)} & \textbf{3.8160} & \textbf{1.9535} & \textbf{1.4255} & \textbf{0.5905} & \textbf{0.3390} & \textbf{0.7500} \\
\underline{TopoBind (MLP)} & \underline{3.8965} & \underline{1.9739} & \underline{1.4499} & \underline{0.5754} & \underline{0.3251} & \underline{0.7214} \\
TopoBind (LASSO) w/o AFF & 3.9254 & 1.9813 & 1.4345 & 0.5783 & 0.3201 & 0.7121 \\
TopoBind (LASSO) w/o Cross-Attention & 4.2049 & 2.0506 & 1.4908 & 0.5339 & 0.2716 & 0.7093 \\
ESM2 (LASSO) & 4.4303 & 2.1042 & 1.5832 & 0.5418 & 0.2896 & 0.6862 \\
\hline
\end{tabular}
\end{table*}

\subsection{Topological Parameter Sensitivity}

We conducted a grid search over two hyperparameters influencing topological feature extraction: the interface contact distance threshold $d$ and the number of top-$k$ persistent homology lifetimes retained. Specifically, $d$ ranged from 6.0 to 8.5 (step 0.5), and $k$ from 3 to 9, balancing geometric richness and computational efficiency.

\begin{figure}[ht]
\centering
\includegraphics[width=0.6\linewidth]{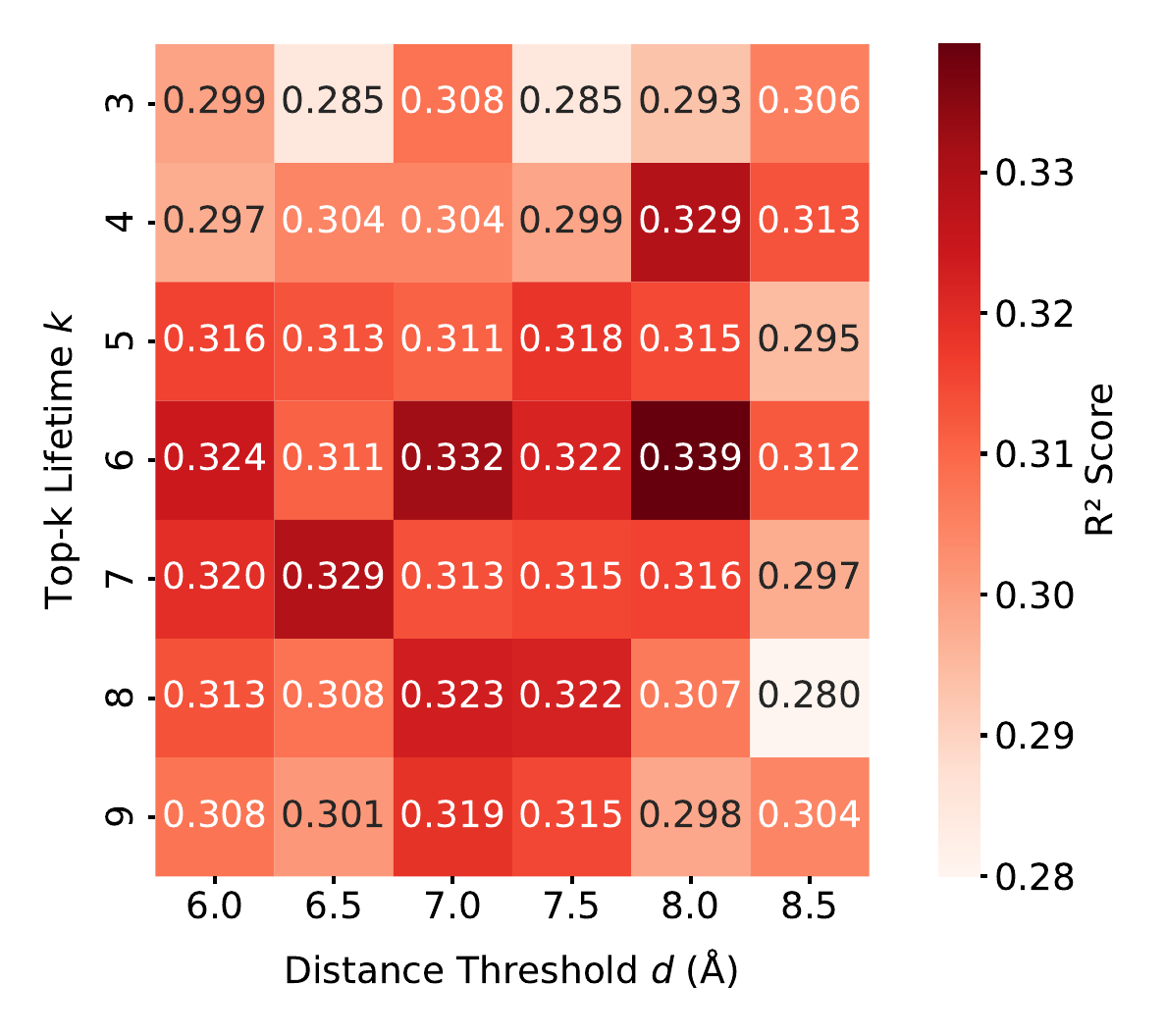}
\caption{$R^2$ score heatmap across different values of interface contact distance threshold $d$ and top-$k$ persistent homology lifetimes $k$. Best performance is observed at $d = 8.0$, $k = 6$.}
\label{fig:hyper_heatmap}
\end{figure}

As shown in Figure~\ref{fig:hyper_heatmap}, the model performs best at $d = 8.0$ and $k = 6$, achieving a peak $R^2$ of 0.339. The surface around this setting is relatively flat, suggesting robustness to small hyperparameter variations. Performance drops when $k$ is too small, likely due to loss of informative topological signals. Similarly, thresholds $d < 7.0$ may miss important contacts, reducing accuracy.

\begin{figure}[htbp]
\centering
\includegraphics[width=0.6\linewidth]{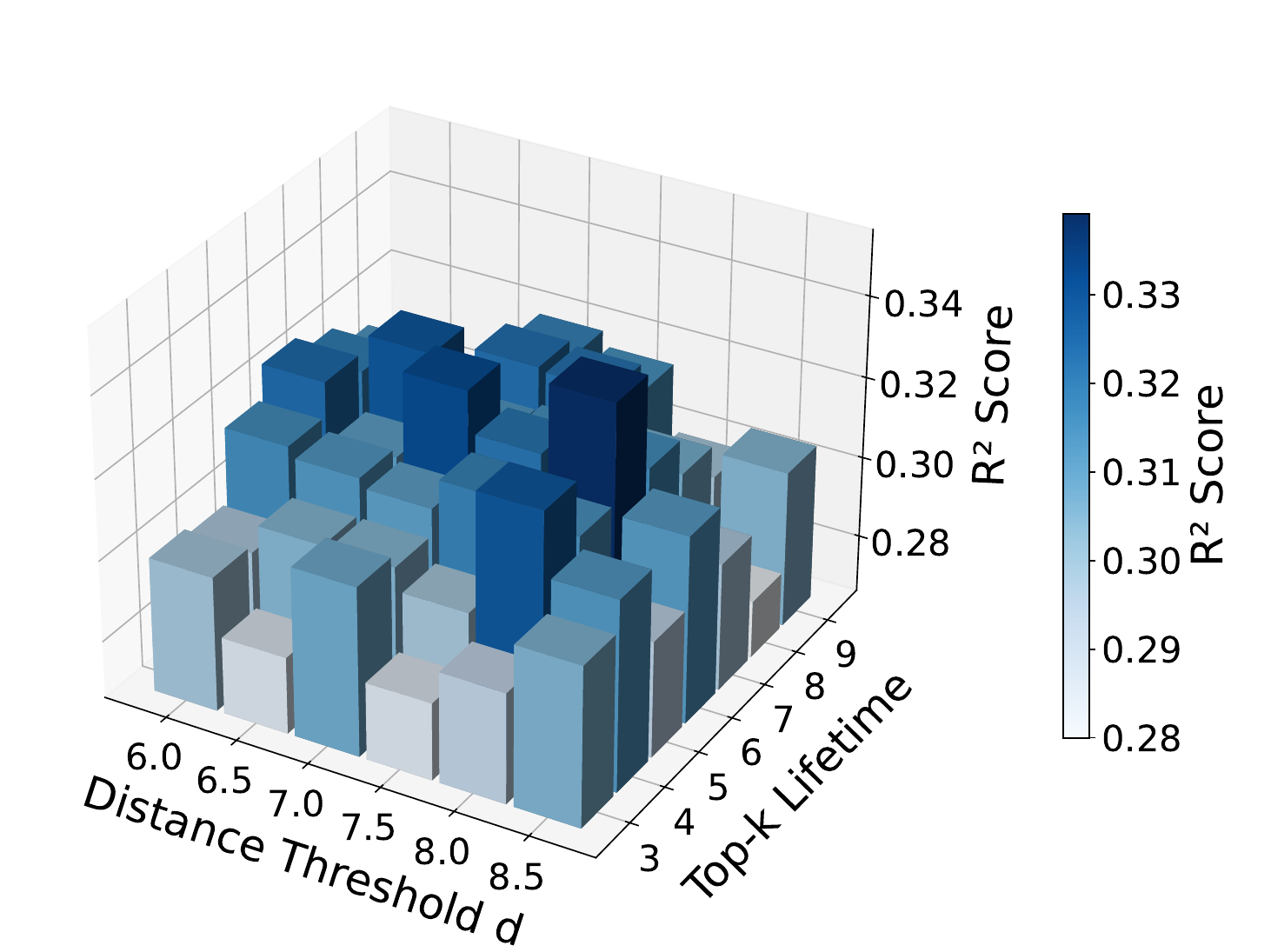}
\caption{3D surface plot showing $R^2$ scores of topology-only models across combinations of interface contact distance threshold $d$ and top-$k$ persistent homology lifetimes $k$. Best performance is observed at $d = 8.0$, $k = 6$.}
\label{fig:parameter-surface}
\end{figure}

Figure~\ref{fig:parameter-surface} shows a 3D surface of $R^2$ across the grid, revealing a non-linear interaction between $d$ and $k$. While the model remains stable across a broad region, performance degrades toward the extremes. In particular, lower $d$ values consistently weaken results, indicating that overly tight thresholds limit structural context. These findings underscore the need to jointly tune both $d$ and $k$ for optimal topological representation.

\section{Conclusion}

We present TopoBind, a multi-modal deep learning framework for predicting antibody-antigen binding free energy ($\Delta G$) by integrating pretrained sequence embeddings with handcrafted structural topology features, including contact maps, interface geometry, Euclidean distances, and persistent homology. The model employs dedicated encoders for each modality, followed by adaptive feature fusion and cross-attention mechanisms to effectively align and integrate complementary information. Experimental results show that TopoBind, when paired with a Lasso regressor, achieves state-of-the-art performance across all regression and classification metrics. Ablation studies further confirm the critical roles of both the adaptive fusion module and sparse linear modeling in enhancing generalization. These findings underscore the importance of incorporating geometric and topological priors in molecular learning, and highlight promising directions such as differentiable topology layers and multi-modal pretraining.

\section*{Acknowledgments}
We thank Dr. Shashvat Shukla (University College London) for insightful discussions.
The computation resources in this study was sponsored by BayVax Biotech Limited. Y.-F. H. thanks support from the Major Program of the National Natural Science Foundation of China (\texttt{92369201}). The research was also supported by the National Key Research and Development Program of China (\texttt{2021YFA0910700}), the Health and Medical Research Fund, the Food and Health Bureau, The Government of the Hong Kong Special Administrative Region (\texttt{COVID1903010},  \texttt{T-11-709/21-N}) to J.-D.H. J.-D.H. also thanks the L \& T Charitable Foundation, the Program for Guangdong Introducing Innovative and Entrepreneurial Teams (\texttt{2019BT02Y198}) and Shenzhen Key Laboratory for Cancer Metastasis and Personalized Therapy (\texttt{ZDSYS20210623091811035}) for their support. 

\bibliographystyle{unsrt}  
\bibliography{references}

\clearpage

\appendix
\section*{Appendix}

\section{Runtime Environment}

All experiments were conducted on a Linux server equipped with high-performance GPU accelerators. The runtime environment is detailed below for reproducibility:

\begin{itemize}
    \item \textbf{OS:} Linux 5.10.0-33-cloud-amd64
    \item \textbf{Python:} 3.10.18
    \item \textbf{CUDA:} 12.1
    \item \textbf{GPU:} NVIDIA A100-SXM4-40GB
    \item \textbf{ESM model:} \texttt{facebook/esm2\_3b\_v1}
    \item \textbf{Main libraries:}
    \begin{itemize}
        \item PyTorch 2.5.1+cu121
        \item NumPy 2.2.6
        \item Scikit-learn 1.7.0
        \item Matplotlib 3.10.3
        \item tqdm 4.67.1
    \end{itemize}
\end{itemize}

All models were trained using PyTorch's mixed-precision training and AdamW optimizer.To ensure reproducibility, we set fixed random seeds for PyTorch and NumPy, and we saved all model checkpoints, predictions, and performance metrics under the \texttt{results/} directory.

\section{Dataset Filtering and Preprocessing Details}

We constructed a curated dataset of antibody-antigen complexes by combining structural and sequence information from public databases. The initial source was a manually annotated CSV file \texttt{alldata.csv}, which included 1705 complex instances annotated with experimental $\Delta G$ values, amino acid sequences of paired antibody/antigen chains, and corresponding PDB IDs with chain identifiers.

To ensure compatibility with both modalities, we identified 303 unique PDB entries that had successfully computed topological features. ESM embeddings were available for 472 entries. The intersection of these two subsets yielded 303 PDB IDs with complete modality coverage. A multi-threaded Python script was used to download the corresponding \texttt{.pdb} files from the RCSB PDB database, and all 303 files were retrieved successfully.

For a small number of entries with missing or incomplete structural data, we attempted to reconstruct atomic coordinates from available sequences using AlphaFold or ColabFold. Entries for which reliable structural reconstruction was not feasible were excluded via automated filtering scripts. The original CSV file was preserved to ensure reproducibility.

It is important to note that each PDB ID in \texttt{alldata.csv} may correspond to multiple chain pairings or complex variants (e.g., multiple antigen chains paired with a fixed antibody). After filtering for valid PDB IDs, all matching complex instances were retained, resulting in a final dataset of 1398 usable samples for training and evaluation. These were split into 978 training, 209 validation, and 211 test samples, with no PDB ID shared across partitions.

This finalized dataset served as the basis for ESM-based sequence embedding extraction and structural topology feature computation.

\section{Topological Feature Engineering}
To capture diverse structural patterns beyond sequence-level representations, we extracted four categories of handcrafted topological features from each antibody-antigen complex: contact statistics, interface geometry, residue-level distances, and persistent homology. These descriptors are designed to encode both local interaction density and global geometric topology. After processing and dimensional alignment, each complex is represented by a unified 100-dimensional topological feature vector, standardized across the dataset.

\subsection{Contact Map Features}
We computed residue-level contact graphs for both antibody and antigen chains. Features include:
\begin{itemize}
  \item \textbf{Contact density:} ratio of residue pairs within 8 \AA;
  \item \textbf{Contact degree statistics:} mean, standard deviation, and maximum of per-residue contact degrees;
  \item \textbf{Clustering coefficient:} triangle-based local contact density.
\end{itemize}
These features aim to describe the overall network connectivity within each chain.

\subsection{Distance Matrix Features}
For both antibody and antigen atoms, we calculated pairwise distances and extracted basic statistical descriptors:
\begin{itemize}
  \item \textbf{Minimum distance}, \textbf{mean distance}, and \textbf{median distance} within each chain.
\end{itemize}

\subsection{Persistent Homology Features}
To characterize the multi-scale topology of the complex, we applied persistent homology up to dimension 2. For each dimension ($H_0$, $H_1$, $H_2$), we computed:
\begin{itemize}
  \item Number of birth-death intervals;
  \item Mean, maximum, sum, and standard deviation of lifetimes;
  \item Top-k longest lifetimes for each dimension.
\end{itemize}
These were separately computed for antibody atoms only.

\subsection{Feature Dimension Summary}
\begin{figure}[ht]
  \centering
  \includegraphics[width=0.45\textwidth]{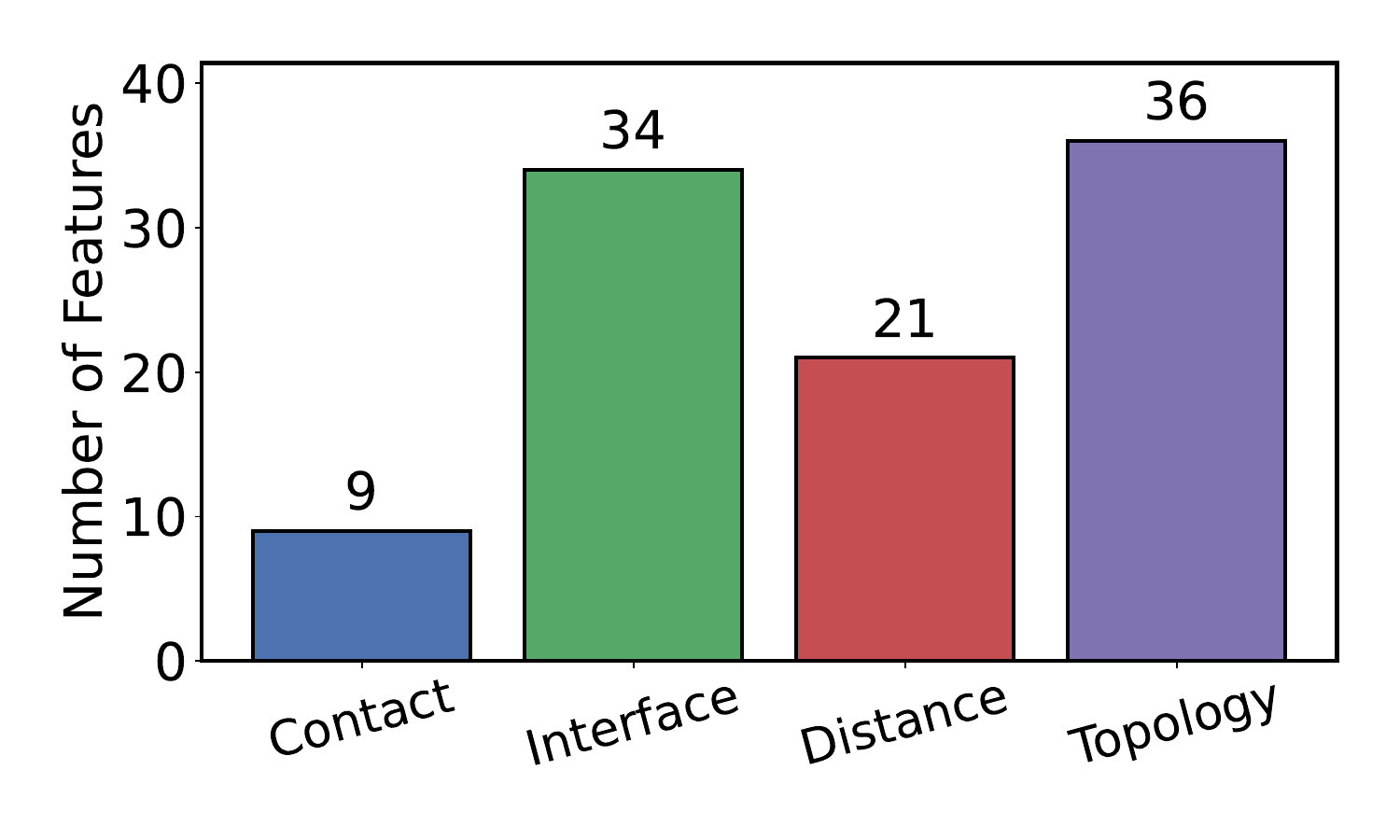}
  \caption{Topological feature distribution by category in TopoBind.}
  \label{fig:topo_feature_counts}
\end{figure}

Figure~\ref{fig:topo_feature_counts} visualizes the composition of the final 100-dimensional topological feature vector. The features are grouped into four semantic categories: contact-level statistics (9 features), interface geometry (34), distance-based descriptors (21), and persistent homology features (36). This categorization reflects the multi-scale nature of antibody-antigen structural interactions.

\section{Feature Fusion and Input Processing}
\label{appendix:fusion}

To support multi-modal learning, we designed a unified processing pipeline that combines pretrained sequence embeddings with handcrafted topological features for each antibody-antigen complex. This section details the input modalities, standardization, and the dimensional alignment procedures used in TopoBind.

\subsection{Input Modalities}

TopoBind takes two types of inputs:

\begin{itemize}
  \item \textbf{Sequence Embeddings:} We use 2560-dimensional embeddings from the ESM-2 pretrained language model, mean-pooled over residue tokens. These embeddings capture evolutionary and contextual information and are stored in \texttt{.pkl} format.
  
  \item \textbf{Topological Features:} Each complex is also represented by a 100-dimensional structural vector, which integrates four categories of handcrafted descriptors: contact statistics, interface geometry, distance-based features, and persistent homology. These are extracted from 3D structural files and stored in \texttt{.npz} format.
\end{itemize}

\subsection{Standardization and Dimensional Alignment}

To ensure consistency and training stability, we apply the following steps:

\begin{itemize}
  \item \textbf{Sample Filtering:} Only complexes with both sequence and topological features are retained, yielding a final dataset of 303 antibody-antigen pairs.
  
  \item \textbf{Padding and Truncation:} All topological vectors are adjusted to a fixed dimension (100) based on the 90th percentile across samples, using zero-padding or truncation as needed.
  
  \item \textbf{Z-score Normalization:} Each modality is normalized independently using the mean and standard deviation computed from the training set.
\end{itemize}

\subsection{Data Loader and Batch Handling}

We implemented a custom PyTorch \texttt{Dataset} class to retrieve aligned ESM and topological features along with the experimental binding free energy  labels. A specialized \texttt{collate\_fn} was used to dynamically pad topological vectors during batching if needed.

Each training instance is a tuple: $(\mathbf{x}_{\text{esm}}, \mathbf{x}_{\text{topo}}, y)$, where $y$ is the true $\Delta G$ value.

\subsection{Cross-Modal Fusion Architecture}

The EnhancedCrossAttentionModel integrates both modalities in a staged fashion:

\begin{itemize}
  \item Each modality is first passed through a two-layer feedforward encoder with LayerNorm, GELU activation, and dropout.
  \item The four branches of topological subfeatures (contact, interface, distance, topology) are adaptively fused via a learnable gating mechanism (AFF).
  \item Two layers of bidirectional cross-attention are applied to align ESM and fused topological encodings, with residual connections, layer normalization, and feedforward blocks at each stage.
  \item The final embeddings are concatenated and used as input for downstream regression via either a multi-layer perceptron or Lasso model.
\end{itemize}

The model is trained to minimize MSE loss using the AdamW optimizer, with learning rate scheduling and early stopping.

\subsection{Overall Pipeline}

The entire workflow, from raw CSV input and structural preprocessing to fused representation learning, regression modeling, and evaluation, is implemented in a modular and reproducible framework. After training, the pipeline automatically outputs evaluation metrics, visualizations, and stores model checkpoints for future inference and analysis.

\section{Hyperparameter Settings and Training Strategy}

We detail the training configuration used for the proposed model in Table~\ref{tab:hyperparams}. All experiments were performed using the AdamW optimizer and mean squared error as the loss function. To ensure training stability and generalization, we applied dropout regularization, learning rate scheduling, and early stopping.

\begin{table}[ht]
\centering
\caption{Hyperparameter settings for model training.}
\label{tab:hyperparams}
\begin{tabular}{ll}
\hline
\textbf{Hyperparameter} & \textbf{Value} \\
\hline
Batch size & 32 \\
Learning rate & 3e-4 \\
Optimizer & AdamW \\
Weight decay & 1e-4 \\
Max epochs & 200 \\
Early stopping patience & 20 \\
Learning rate scheduler & ReduceLROnPlateau \\
Dropout rate & 0.1 \\
Hidden dimension & 256 \\
Number of attention heads & 8 \\
Number of CrossAttention layers & 2 \\
\hline
\end{tabular}
\end{table}

\subsection{Training Procedure}

The dataset was split into training, validation, and test sets using a 70\%/15\%/15\% ratio with a fixed random seed (\texttt{torch.manual\_seed(42)}) for reproducibility. We employed mixed precision training (AMP) to accelerate convergence and reduce memory usage on compatible GPUs.

The model was trained using early stopping based on validation loss, with learning rate decay triggered by a plateau scheduler. At each epoch, both training and validation losses were logged. After training, we evaluated the model on the held-out test set using six metrics: MSE, RMSE, MAE, Pearson correlation, $R^2$, and classification accuracy .

Model checkpoints, evaluation results, and training logs are saved automatically to facilitate reproducibility and downstream inference.

\subsection{Parameter Sensitivity Analysis}
To investigate the impact of key hyperparameters on model performance, we conducted controlled experiments by varying the filtration threshold in topological feature extraction and the number of top persistent lifetimes retained. For each configuration, we retrained the model and reported its predictive performance.

To illustrate the results under the optimal setting ($d=8.0$\AA, $k=6$), we present the training loss curve, the predicted vs. true $\Delta G$ values, and the error distribution. These results highlight the convergence stability and predictive consistency of our model under this parameter configuration.

\begin{figure}[ht]
    \centering
    \includegraphics[width=0.6\linewidth]{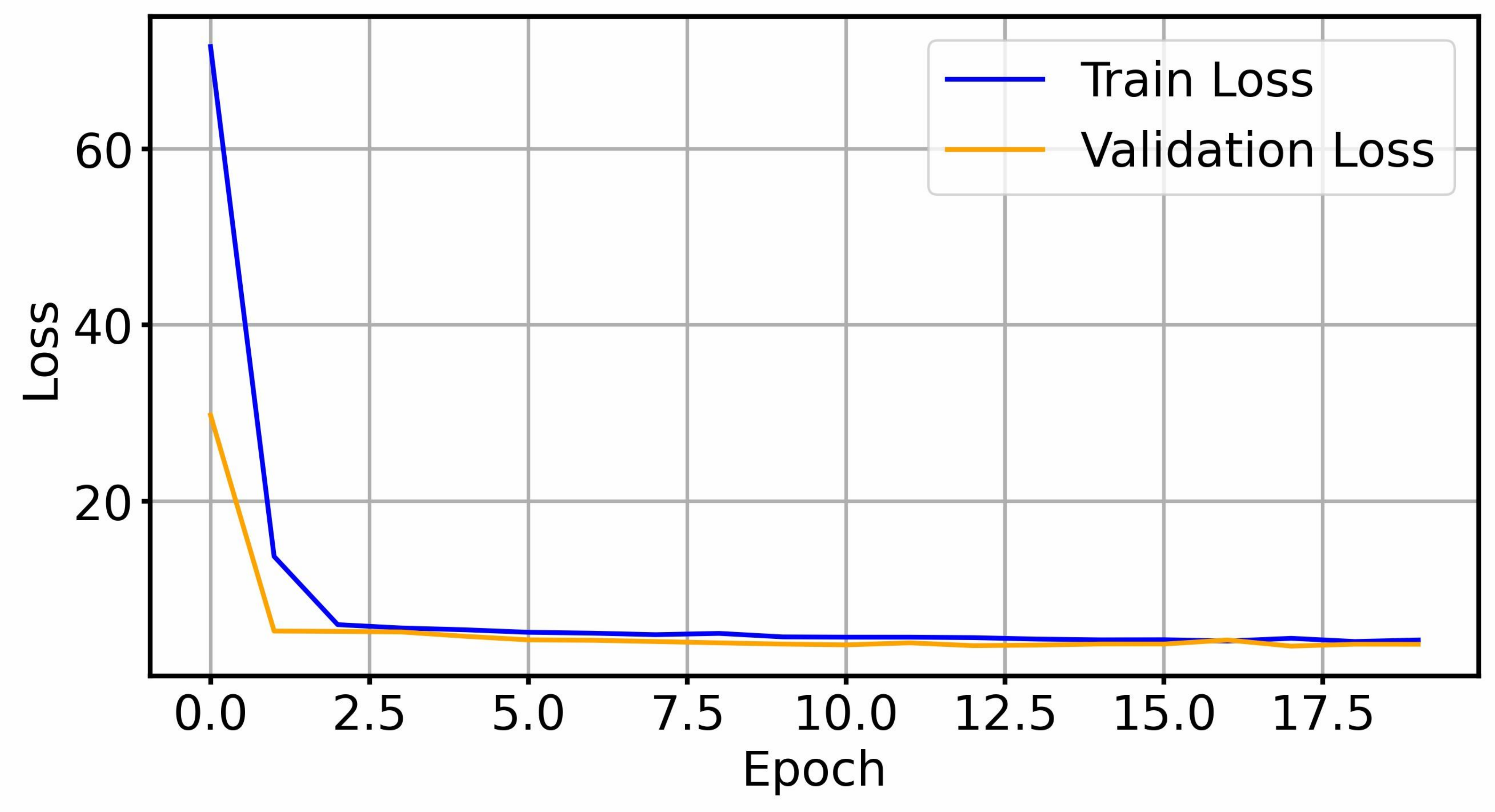}
    \caption{Training and validation loss under the optimal setting.}
\end{figure}

\begin{figure}[ht]
    \centering
    \includegraphics[width=0.6\linewidth]{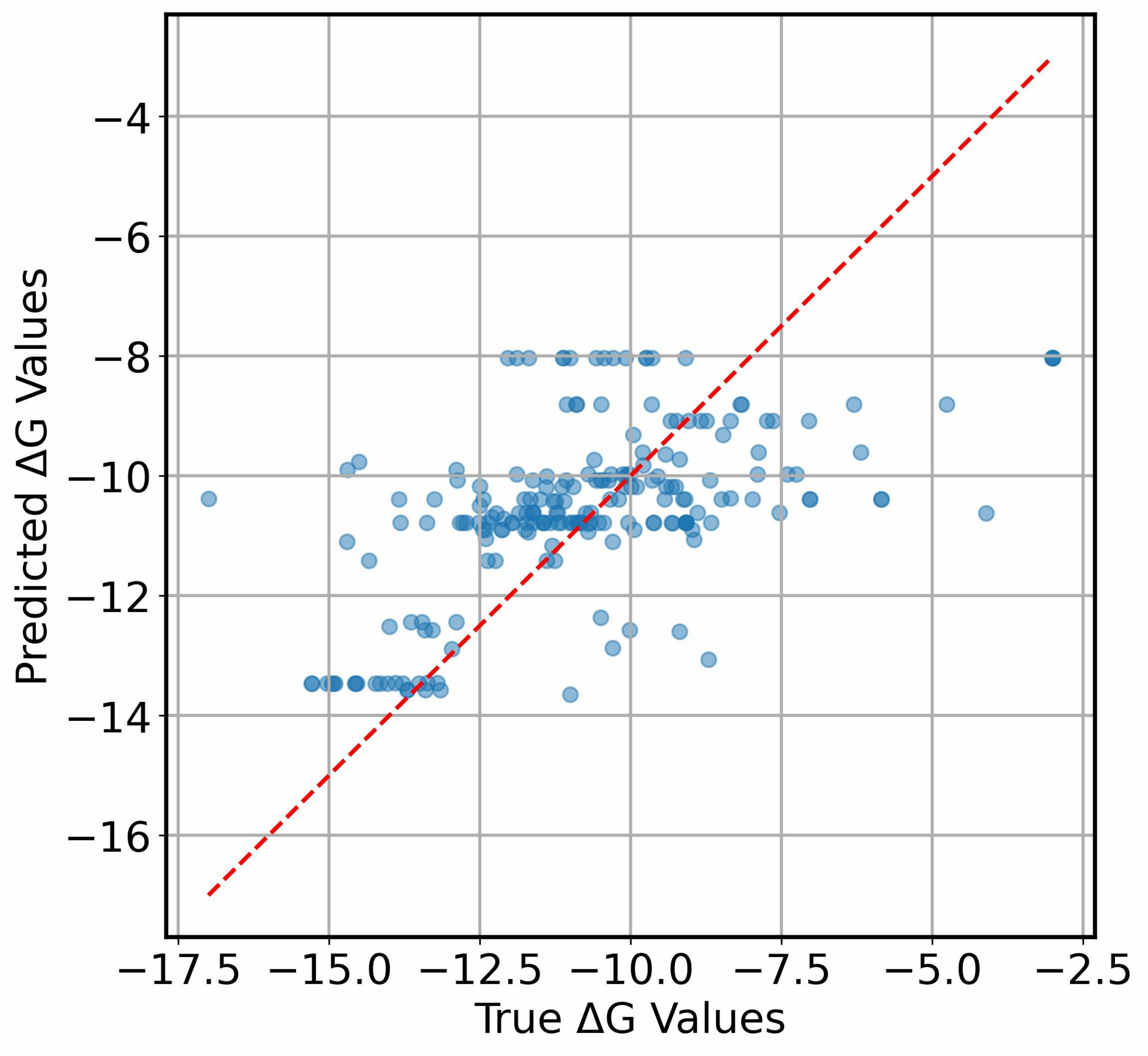}
    \caption{Predicted vs. true $\Delta G$ values under the optimal setting.}
\end{figure}

\begin{figure}[ht]
    \centering
    \includegraphics[width=0.6\linewidth]{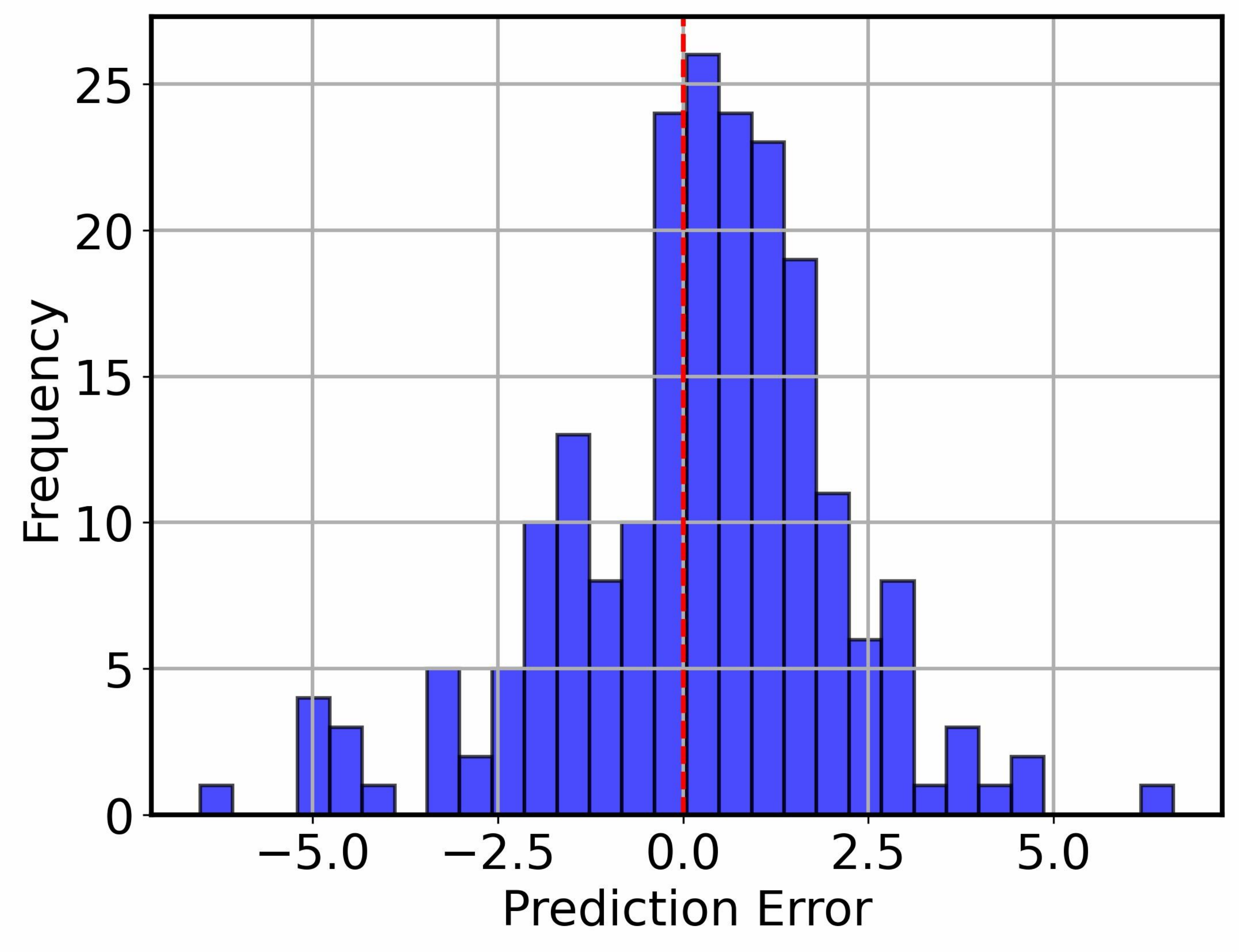}
    \caption{Distribution of prediction error under the optimal setting.}
\end{figure}

\end{document}